\documentclass[runningheads]{llncs}
\usepackage[T1]{fontenc}
\usepackage[latin9]{inputenc}
\usepackage{xcolor}
\usepackage{textcomp}
\usepackage{mathtools}
\usepackage{url}
\usepackage{amstext}
\usepackage{amsmath}
\usepackage{graphicx}
\usepackage{comment}
\usepackage{epstopdf}
\usepackage{subfig}
\definecolor{britishracinggreen}{rgb}{0.05, 0.5, 0.0}
\definecolor{magenta}{rgb}{1.0, 0.0, 1.0}
\emergencystretch=\maxdimen
\makeatletter

\newcommand{\lyxmathsym}[1]{\ifmmode\begingroup\def\b@ld{bold}
  \text{\ifx\math@version\b@ld\bfseries\fi#1}\endgroup\else#1\fi}

\begin{document}
\title{Computational modelling of complex multiphase behavior of environmentally-friendly materials for sustainable technological solutions}
\titlerunning{Computational modelling of complex multiphase behavior}
%
\author{Akshayveer\inst{1}\orcidID{0000-0003-0289-5357} \and
Federico C. Buroni\inst{2}\orcidID{0000-0001-9746-9478} \and \\
Roderick Melnik\inst{1}\orcidID{0000-0002-1560-6684}\and \\
Luis Rodriguez-Tembleque\inst{3}\orcidID{0000-0003-2993-8361} \and\\
Andres Saez\inst{3}\orcidID{0000-0001-5734-6238}}
\authorrunning{Akshayveer et al.}
%
\institute{MS2Discovery Interdisciplinary Research Institute, Wilfrid Laurier
University, Waterloo, Ontario N2L3C5, Canada\\ 
\email{\{aakshayveer,rmelnik\}@wlu.ca} \and
Department of Mechanical Engineering and Manufacturing, Universidad
de Sevilla, Camino de los Descubrimientos s/n, Seville E-41092, Spain
\\
\email{fburoni@us.es} \and
Department of Continuum Mechanics and Structural Analysis, Universidad
de Sevilla, Camino de los Descubrimientos s/n, Seville E-41092, Spain\\
\email{\{luisroteso,andres\}@us.es}}
\maketitle              
\vspace{-0.50cm}
\begin{abstract}
This study presents a computational framework to investigate and predict the complicated multiphase properties of eco-friendly lead-free piezoelectric materials, which are crucial for sustainable technological progress. Although their electromechanical properties vary by phase, lead-free piezoelectric materials show a considerable thermo-electromechanical response. Lead-free materials such as Bi$_{0.5}$Na$_{0,5}$TiO$_{3}$ (BNT) and other BNT-type piezoelectric materials transition to rhombohedral (R3c), orthorhombic (Pnma), tetragonal (P4bm), and cubic (Cc) phases with temperature variation. These phases are determined by the symmetry and alignment of the ferroelectric domains. Multiple phases can occur simultaneously under specific thermal, electrical, and mechanical conditions, leading in complex multiphase behaviour. These materials' performance must be assessed by studying such behaviour. This study uses Landau-Ginzburg-Devonshire theory to simulate material micro-domain phase transitions. The computational model for BNT-type piezoelectric material covers temperature-induced ferroelectric domain switching and phase transitions. Therefore, the developed computational approach will assist us in better understanding the influence of these materials' complex multiphase behaviour on creating sustainable solutions with green technologies.
\vspace{-0.250cm}
\end{abstract}
\vspace{-0.50cm}
\section{Introduction}
Piezoelectric materials transform electrical energy into mechanical energy, essential in electronic devices like sensors, actuators, transducers, and energy harvesters \cite{JAFFE1958}. They also have features like energy storage, field-induced strain, pyroelectricity, and polarization switching, leading to their use in dielectric capacitors, ferroelectric memory, and infrared detectors \cite{Maurya2018}. Lead-based ceramics (such as Pb(Zr,Ti)O$_{3}$ (PZT))  are commonly used due to their exceptional piezoelectric characteristics (piezoelectric coefficient $d_{33}$ of 600-700 $pC/N$, high Curie temperature $T_{c}$ of around 200 \textcelsius{}, and high strain responsiveness of 0.8\%) \cite{Dutta2011}. However, lead oxide is hazardous, leading to the need for lead-free alternatives such as barium titanate (BaTiO$_{3}$, BT), potassium sodium niobite (K$_{0.5}$Na$_{0.5}$NbO$_{3}$, KNN), and bismuth sodium titanate (Bi$_{0.5}$Na$_{0.5}$TiO$_{3}$, BNT) \cite{Wang2021}.

Moreover, KNN and BT-based piezoelectric materials show improved response ($d_{33}\sim500pc/N$) and ($d_{33}\sim445\pm20pc/N$), however; the low Curie temperature ($T_{c}$) for KNN-based composites ($T_{c}\sim200\lyxmathsym{\textcelsius}$) and BT-based composites ($T_{c}<100\lyxmathsym{\textcelsius}$) restricts their use for high temperature applications \cite{Wang2021}. Pure BNT exhibited moderate piezoelectric characteristics ($d_{33}<100pc/N$) and a strong coercive electric field ($E_{c}\sim70kV/cm$) \cite{Hao2019}. BNT's high $T_{c}$ (320\textcelsius) \cite{Kumari2022} improves its suitability for high-temperature haptic applications. Additionally, BNT-based composites show higher strain than KNN and BT-based composites at temperatures above 200\textcelsius{ } \cite{Zhou2021}, making them ideal for high-temperature actuators.

BNT has complex phase structure. The structure can be R3c, monoclinic (Cc), or a mix of both below depolarization temperature ($T_{d}\sim$200\textcelsius) \cite{Rao2013,Jones2002}, depending on thermal, electrical, and mechanical treatments. BNT, with its nonergodic relaxor (NR) and ferroelectric (FE) properties, exhibits a square polarization-electric field loop, high remnant polarization $P_{r}$, and distinct macro-piezoelectricity $d_{33}$ from room temperature to dipolar freezing temperature $T_{f}$ (190\textcelsius) \cite{Kreisel2002}, making it suitable for high-temperature haptic applications. Higher temperatures than $T_{f}$ cause the BNT material to transition from antiferroelectric (AFE) to ferroelectric (FE) phase, reducing its piezoelectric properties. Beyond $T_{d}$, the material exhibits ergodic relaxor (ER) behaviour and becomes fully AFE. Several studies \cite{Dorcet2009} identified an intermediate Pnma phase with AFE properties between 200-320\textcelsius. However, most researchers now consider it a non-polar or weakly-polar phase \cite{Rao2013}, making it more suitable for actuator applications due to its high starin response. At temperatures above 320\textcelsius, BNT phase transition becomes more complicated \cite{Jones2002}. BNT enters P4bm, a tetragonal symmetric paraelectric phase, at 320\textcelsius. The P4bm symmetry shows octahedral tilting and anti-parallel displacement of Na$^{+}$/Bi$^{3+}$ and Ti$^{4+}$ cations along [001]P$_{C}$, with just 0.2 percent cubic structural deformation \cite{Jones2002}. At temperatures over 520 \textcelsius, the crystal structure becomes cubic Pm3m. A transition between cubic and tetragonal ferroelectric phases or between cubic and super-paraelectric phases may occur around 520 \textcelsius{}. The BNT exhibits non-negligible hysteresis and $P_{r}$ at high temperatures (320-520\textcelsius) \cite{Li2019}, making it suitable for sensor and haptic applications. Low $P_{r}$ values in the BNT's paraelectric phase over 520\textcelsius{} restrict sensing and haptic applications.

The forgoing research shows that BNT exhibits difficult multi-phase co-existence over different temperature regimes, affecting strain response, polarization, and energy storage density. BNT's elastic, piezoelectric, and energy storage characteristics make it useful in various regimes. Novel uses of this material need a careful investigation of its complicated phase change behaviour at different temperatures. Under varied thermal, mechanical, and electrical boundary conditions, we studied BNT material phase transition changes using an experimentally validated computational model. Considering Landau-Ginzburg-Devonshire free energy in the computational model allows BNT material micro-domain switching during phase development and transitions at different temperatures. Although BNT-type piezoelectric materials have a convoluted phase structure, our computational method records correctly micro-domain switching and phase transitions under various heat settings. Thus, this research will help us better understand complicated materials' behaviours, leading to sustainable and eco-friendly sensors, actuators, haptic applications, and human-computer interfaces.
\vspace{-0.25cm}
\section{Methodology and model development}
A two-dimensional piezoelectric composite design with micro-scale piezoelectric inclusions (BNT) is examined for transient thermo-electromechanical behavior with complicated phase shift and domain switching. The following sub-sections will discuss the composite architecture, the coupled thermo-piezoelectric model used to study the composite's behavior,
the materials models that govern its dielectric and mechanical properties, and the boundary conditions used to compute specific effective electro-elastic coefficients of interest. 
\vspace{-0.25cm}
\subsection{Geometrical description and boundary conditions}
\begin{figure}[!t]
\vspace{-0.5cm}
\centering
\includegraphics[scale=0.23]{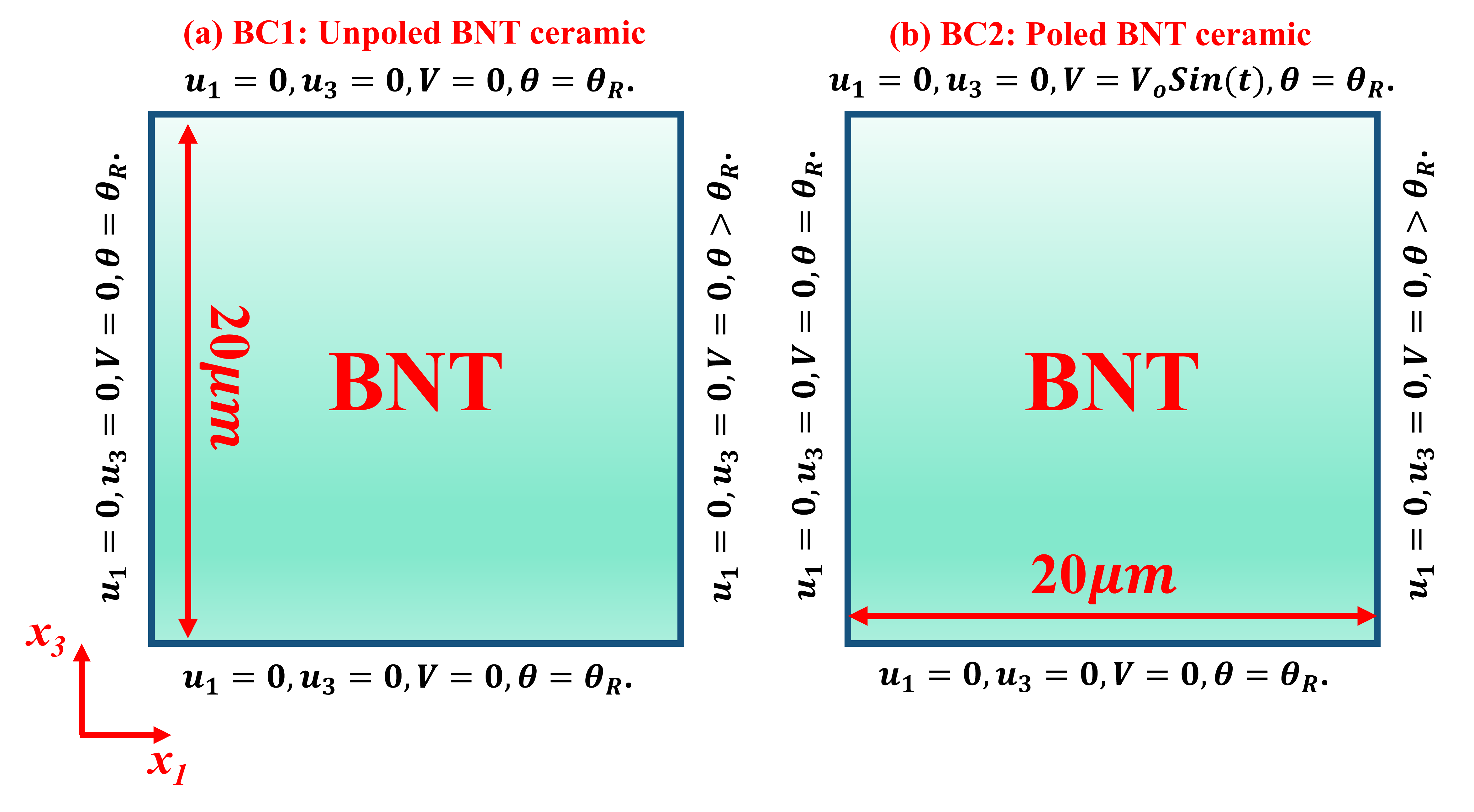}
\caption{Schematic showing the geometric description and boundary conditions of BNT ceramic.\label{fig:Schematic1}}
\vspace{-0.5cm}
\end{figure}

In Figure \ref{fig:Schematic1}, the piezoelectric ceramic is shown as a two-dimensional RVE in the $x_{1}-x_{3}$ plane. A 20 $\mu m$ square BNT composite ceramic is used. We investigate the composite material's P-E hysteresis curve to assess its piezoelectric and ferroelectric phase transition capabilities. Two boundary conditions, BC1 and BC2, are needed to derive these curves, as shown in Fig. \ref{fig:Schematic1}(a) and (b).
The reference temperature $\theta_{R}$ is the ambient temperature (27\textcelsius), whereas the thermal boundary conditions at the right wall can reach up to 520\textcelsius.
The electric potential in BC2 follows a sine function of time (t) to provide P-E curves under certain thermal boundary conditions for polarized and depolarized BNT composites. Initial conditions are critical for transient phase-field thermo-electromechanical modelling of BNT-based piezoelectric composite, together with boundary conditions. Initial conditions are
$u_{i}(t=0)=0,~i=1, 3; ~V(t=0)=0, ~ \theta(t=0)=\theta_{R}.$

\subsection{Mathematical model}
 We introduce the thermo-electromechanical
model used to examine the BNT ceramic design described in Section 2.1
and shown in Fig. \ref{fig:Schematic1}. The model examines phase
transition and ferroelectric domain transitions by incorporating a
Landau-Ginzburg-Devonshire free energy (e.g. in \cite{Ahluwalia2016} and Table \ref{tab:Eqns}) in the Helmholtz free energy of the system.
 The constants are denoted as $\pmb{C}$ for elastic, $\pmb{\epsilon}$ for dielectric, $\pmb{e}$ for piezoelectric, $\pmb{\mu}$ for flexoelectric, $\pmb{\beta}$ for thermal expansion coefficient, and $\pmb{\eta}$ for thermoelectric coefficient. Additionaly, $\pmb{\overrightarrow{E}}$, $\pmb{\varepsilon(\overrightarrow{u}})$, $\pmb{\varepsilon}^{el}$, $\pmb{\varepsilon^{t}(\overrightarrow{p})}$, and $\theta$ represent the electric field intensity vector, total mechanical strain tensor,  elastic strain tensor, the transformation
strain tensor, and temperature respectively.
\begin{table}[!t]
\vspace{-0.50cm}
\caption{\label{tab:Eqns}Equations for computational model.}
\centering
\begin{tabular}{@{}l}
\hline 
\textbf{Total free energy of the system:} \\
$\phi(\pmb{\overrightarrow{E}},\pmb{\varepsilon}(\pmb{\overrightarrow{u}}),\nabla\pmb{\varepsilon}, \pmb{\overrightarrow{p}},\pmb{\nabla\overrightarrow{p}},\theta) = \frac{1}{2}\lambda\left|\pmb{\nabla\overrightarrow{p}}\right|^{2}+W(\theta,\pmb{\overrightarrow{p}})+\frac{1}{2}\pmb{\varepsilon^{el}C\varepsilon^{el}}-\pmb{\mu\overrightarrow{E}\nabla\varepsilon^{el}}-\frac{1}{2}\pmb{\epsilon{\overrightarrow{E}}^{2}}$\\
$~~~~~~~~~~~~~~~~~~~~~~~~~~~~~~~~~~~~~~-(\theta-\theta_{R})\pmb{\beta\varepsilon}(\pmb{\overrightarrow{u}})-\pmb{e\overrightarrow{E}\varepsilon^{el}}-\pmb{\overrightarrow{p}\overrightarrow{E}}-\pmb{\eta}(\theta-\theta_{R})\pmb{\overrightarrow{E}},$\\
\textbf{Strain tensors:} \\
$\pmb{\varepsilon}(\pmb{\overrightarrow{u}})=  \frac{1}{2}(\pmb{\nabla\overrightarrow{u}+\nabla \overrightarrow{u}}^T)=\pmb{\varepsilon^{el}}+\pmb{\varepsilon^{t}}(\pmb{\overrightarrow{p}}),$ 
$\pmb{\varepsilon^{t}}(\pmb{\overrightarrow{p}})  =  \gamma\left|\pmb{\overrightarrow{p}}\right|(\pmb{\overrightarrow{n}}\otimes\pmb{\overrightarrow{n}}-\frac{1}{3}\pmb{I}),\pmb{\overrightarrow{n}}\coloneqq\frac{\pmb{\overrightarrow{p}}}{\left|\pmb{\overrightarrow{p}}\right|},$\\
\textbf{Landau-Ginzburg-Devonshire free energy  function:} \\
$W(\theta,\pmb{\overrightarrow{p}})  = \alpha_{1}\frac{\theta_{c}-\theta}{\theta_{c}}(p_{1}^{2}+p_{2}^{2}+p_{3}^{2})+\alpha_{11}(p_{1}^{4}+p_{2}^{4}+p_{3}^{4})+\alpha_{12}\frac{\theta_{c}-\theta}{\theta_{c}}(p_{1}^{2}p_{2}^{2}+p_{2}^{2}p_{3}^{2}+ p_{3}^{2}p_{1}^{2})$\\ $~~~~~~~~~~~~~~~+\alpha_{111}(p_{1}^{6}+p_{2}^{6}+p_{3}^{6})$.\\
\textbf{Strategies of minimizing $W(\theta,\pmb{\overrightarrow{p}})$ for zero electric field at $\theta<\theta_{0}$:}\\
Cubic phase: $p_{1}=p_{2}=p_{3}=0,$ \\Tetragonal phase: $p_{1}=p_{2}=0,~ \alpha_{1}\frac{\theta_{c}-\theta}{\theta_{c}}+2\alpha_{11}p_{3}^{2}+3\alpha_{111}p_{3}^{4}=0,$\\
Orthorombic phase:$p_{1}=0, p_{2}=p_{3} \quad \alpha_{1}\frac{\theta_{c}-\theta}{\theta_{c}}+(2\alpha_{11}+\alpha_{12}\frac{\theta_{c}-\theta}{\theta_{c}})p_{3}^{2}+3\alpha_{111}p_{3}^{4}=0,$\\
Rhombohedral phase: $p_{1}=p_{2}=p_{3} \quad \alpha_{1}\frac{\theta_{c}-\theta}{\theta_{c}}+2(\alpha_{11}+\alpha_{12}\frac{\theta_{c}-\theta}{\theta_{c}})p_{3}^{2}+3\alpha_{111}p_{3}^{4}=0.$\\
\textbf{$W(\theta,\pmb{\overrightarrow{p}})$ for different phases:}\\
Cubic phase: $W(\theta,\pmb{\overrightarrow{p}})=0$, Tetragonal phase:  $W(\theta,\pmb{\overrightarrow{p}})=\alpha_{1}\frac{\theta_{c}-\theta}{\theta_{c}}p_{3}^{2}+\alpha_{11}p_{3}^{4}+\alpha_{111}p_{3}^{6},$\\
Orthorombic phase:  $W(\theta,\pmb{\overrightarrow{p}})=2\alpha_{1}\frac{\theta_{c}-\theta}{\theta_{c}}p_{3}^{2}+(2\alpha_{11}+\alpha_{12}\frac{\theta_{c}-\theta}{\theta_{c}})p_{3}^{4}+2\alpha_{111}p_{3}^{6},$\\
Rhombohedral phase:  $W(\theta,\pmb{\overrightarrow{p}})=3\alpha_{1}\frac{\theta_{c}-\theta}{\theta_{c}}p_{3}^{2}+3(\alpha_{11}+\alpha_{12}\frac{\theta_{c}-\theta}{\theta_{c}})p_{3}^{4}+3\alpha_{111}p_{3}^{6},$\\
\textbf{$W(\theta,\pmb{\overrightarrow{p}})$ in proximity of Curie temperature $\theta_{c}$:}\\
$W(\theta,\pmb{\overrightarrow{p}})  = \alpha_{1c}p_{3c}^{2}+\alpha_{11}p_{3c}^{4}+\alpha_{111}p_{3c}^{6}= 0$\\
\textbf{The values of $\alpha_{1}$, $\alpha_{11}$, $\alpha_{12}$, and $\alpha_{111}$:}\\
$2\alpha_{1} = \frac{1}{\chi}=\frac{(\theta-\theta_{0})}{(\theta_{c}-\theta_{0})}\alpha_{1c},$ $\alpha_{11} = \frac{-2\alpha_{1c}}{p_{3c}^{2}},\quad \alpha_{12}=-a\alpha{11}, \quad \alpha_{111}  =  \frac{\alpha_{1c}}{p_{3c}^{4}}.$\\
\textbf{Constitutive equations:}\\
$\phi_{\pmb{\varepsilon}}=\pmb{\sigma}=\pmb{C}(\pmb{\varepsilon}(\pmb{\overrightarrow{u}})-\pmb{\varepsilon^{t}}(\pmb{\overrightarrow{p}}))-\pmb{e}\pmb{\overrightarrow{E}}-\pmb{\beta}(\theta-\theta_{R}),$ $\phi_{\pmb{\nabla\varepsilon}} = \pmb{\hat{\sigma}}=\pmb{\mu}\pmb{\overrightarrow{E}},$\\
$-\phi_{\pmb{\overrightarrow{E}}} =  \pmb{\overrightarrow{D}}=\pmb{\epsilon}\pmb{\overrightarrow{E}}+\pmb{e}(\pmb{\varepsilon}(\pmb{\overrightarrow{u}})-\pmb{\varepsilon^{t}}(\pmb{\overrightarrow{p}}))+\pmb{\eta}(\theta-\theta_{R})+\pmb{\mu}(\pmb{\nabla\varepsilon}(\pmb{\overrightarrow{u}})-\pmb{\nabla\varepsilon^{t}}(\pmb{\overrightarrow{p}}))+\pmb{\overrightarrow{p}},$\\$-\phi_{\theta} = S=\pmb{\beta}\pmb{\varepsilon}(\pmb{\overrightarrow{u}})-\pmb{\eta}\pmb{\overrightarrow{E}}+W_{\theta}(\theta,\pmb{\overrightarrow{p}}),$
$\phi_{\pmb{\overrightarrow{p}}}=W_{\theta,\pmb{\overrightarrow{p}}}(\pmb{\overrightarrow{p}}),\pmb{\overrightarrow{q}}=-\pmb{k}\nabla\theta,\pmb{\overrightarrow{E}}=-\nabla V,\pmb{k}>0.$\\
\textbf{Governing equations:}\\
$\rho\ddot{\pmb{\overrightarrow{u}}} =  \nabla\cdotp(\pmb{\sigma}-\pmb{\hat{\sigma}})+\pmb{\overrightarrow{F}},$$\tau\dot{\pmb{\overrightarrow{p}}} =  \nabla\cdotp(\lambda\nabla\pmb{\overrightarrow{p}})-\phi_{\pmb{\overrightarrow{p}}},$$-\theta\dot{{\phi_{\theta}}} = \tau\dot{\pmb{\overrightarrow{p}}}^{2}+\nabla\cdotp(\pmb{k}\nabla\theta),$$\nabla\cdotp \pmb{\overrightarrow{D}} = \rho_{e},$\\
$\rho$: mass density,$\rho_{e}$: electric charge density,$k$: thermal conductivity,$\pmb{\overrightarrow{F}}$: external force,\\$\chi$: dielectric susceptibility,$\tau$: inverse mobility coefficient,$\lambda$: Interface energy coefficient.\\
\hline
\end{tabular}
\vspace{-0.5cm}
\end{table}
In Table \ref{tab:Eqns}, we have collected all governing equations and constitutive relationships for the computational domain and boundary conditions, as shown in Fig. \ref{fig:Schematic1}.
\vspace{-0.25cm}
\subsection{Material properties and computational implementation}
Microscale BNT piezoelectric ceramic was chosen for its ability to regulate grain size and create excellent polycrystallinity and piezoelectric response. The temperature dependent material characteristics of BNT ceramic at the specified reference temperature ($\theta_{R}$) are adopted from Hiruma et al. \cite{Hiruma2009}. 

All phenomenological relations are discretized for the computational domain and related to the governing equations. The computational region was partitioned into a reasonable number of varied triangular mesh components using mesh convergence analysis. The residual polarization $P_{r}$ was analyzed at 75\textcelsius\ for various grid counts. Results indicate that 15625 grids result in little change (less than 0.5\%) in $P_{r}$ values, indicating that this is the optimal number. Grid layouts with mesh sizes between 0.16$\mu m$ and 10$nm$ are employed in this study. The discretized equations are solved using finite elements in the computational domain with BNT material boundary conditions. Modelling phase transitions at specific temperatures is challenging due to varying polarization vector orientations (see Table  \ref{tab:Eqns}). The complicated coupling of phase transitions was addressed by selective scale switching of polarization vector coefficients.
\section{Results and discussions}
\vspace{-0.25cm}
Lead-free piezoelectric materials like BNT are versatile for ecologically friendly sustainable technology due to their complex multiphase properties. At various temperatures, BNT exhibits complex ferroelectric domain switching, resulting in multiple phases and complex multiphase states. BNT material's polarization values and strain response vary with temperature, making it suited for sensor, actuator, and haptic device application in computer-human interactions across multiple temperatures. We provide some outstanding scenarios of BNT's P-E characteristics curve analysis to establish its potential as a multifunctional eco-friendly material. Prominent P-E features imply BNT's sensor and haptic compatibility.
\vspace{-0.25cm}
\subsection{Experimental validation of the developed computational model}
\begin{figure}[h]
\vspace{-0.5cm}
\centering
\subfloat{\includegraphics[scale=0.23]{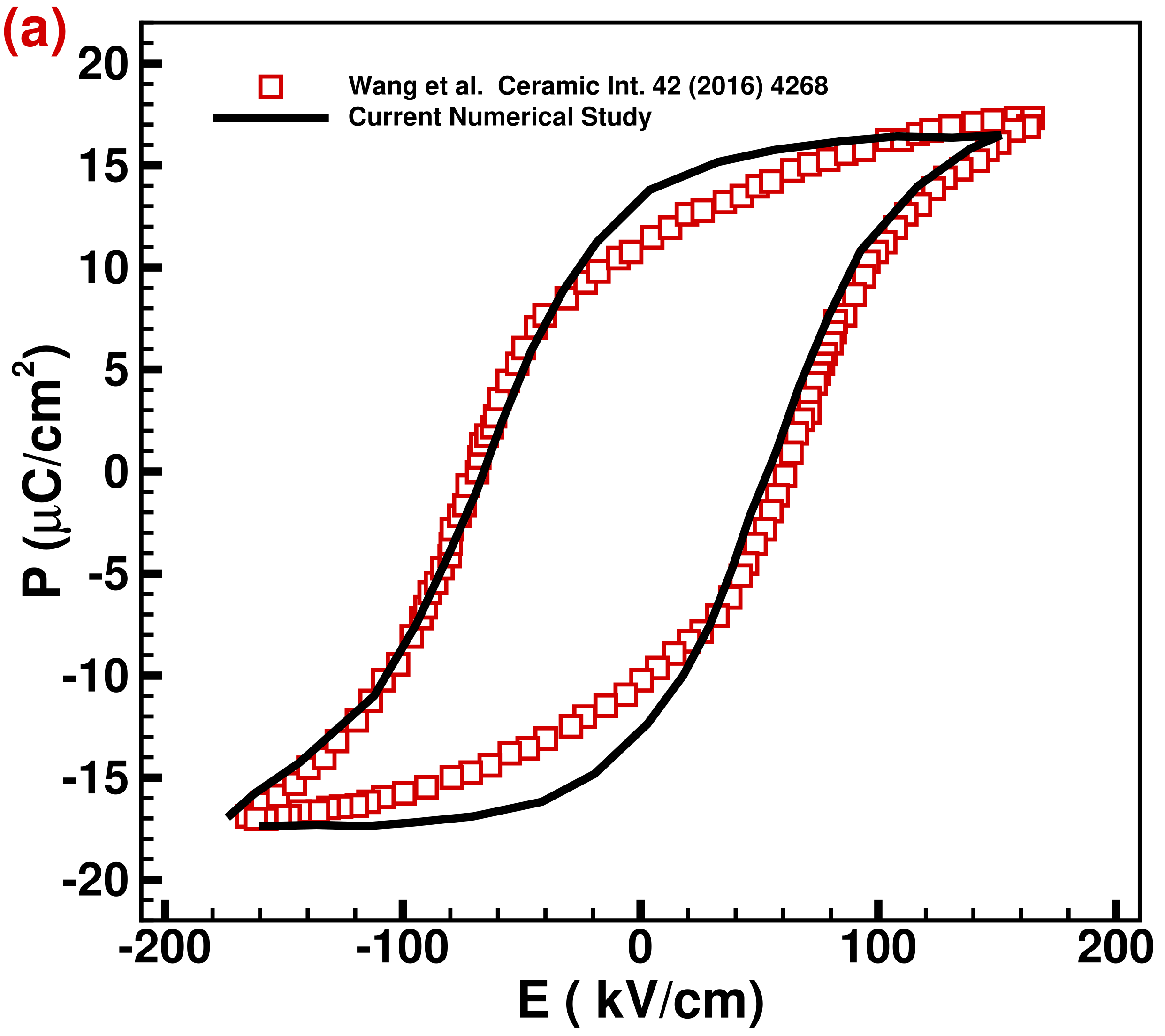}}
\subfloat{\includegraphics[scale=0.23]{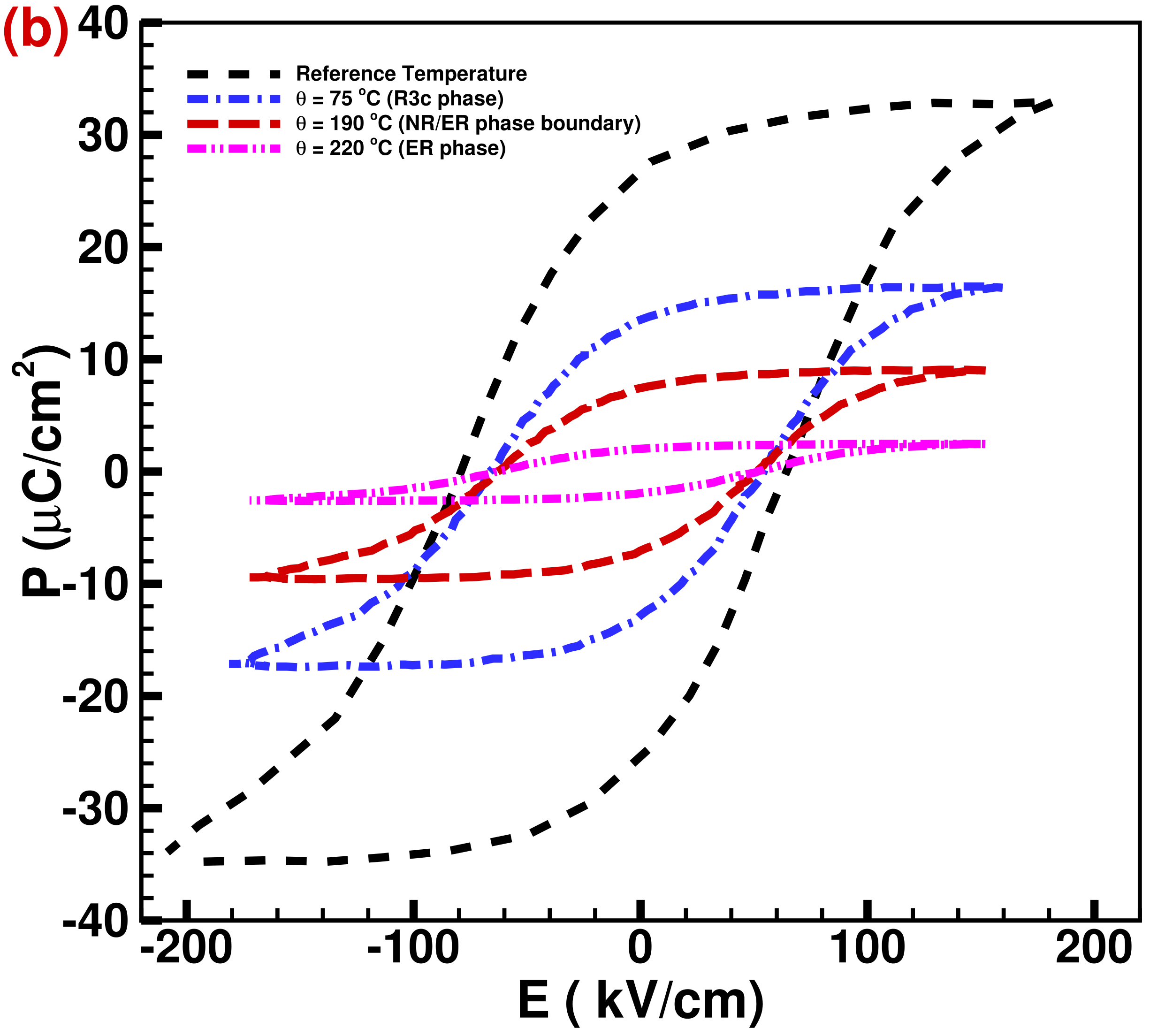}}
\caption{(a) Comparison of P-E hysteresis curve of computational model to experimental
work \cite{Wang2016}, and (b) P-E hysteresis curves for BNT ceramics at different temperatures and phase regimes.}
\label{fig:Comparison-of-P-E}
\vspace{-0.75cm}
\end{figure}

To validate our computational models, we reproduced the polarization hysteresis loops at 1Hz with an external electric field of 160 $kV/cm$ from the experimental study in \cite{Wang2016}. Our computational model replicates the experimental results in Fig. \ref{fig:Comparison-of-P-E}(a) at 75\textcelsius (R3c phase). The numerical model's P-E curve closely matches the experimental curve presented in \cite{Wang2016}. Furthermore, the computational model's ability to closely approximate residual polarization and coercive electric field values, differing by less than 1\% (as it is not possible to replicate experimental circumstances fully), underscores its potential for studying BNT's phase transition performance and its implications on various applications under diverse physical and environmental conditions. The computational model established here can be used for future investigations on the complicated behaviour of BNT and BNT-based composites. 
\vspace{-0.25cm}
\subsection{Complex nonlinear P-E characteristics curves in different phase regimes}
BNT ceramics, a lead-free and eco-friendly material, display varying polarization properties at different temperatures. In particular, BNT possesses an R3c phase structure before depolarization, although remnant polarization should diminish with temperature. The investigation was conducted at 1 Hz, 160 kV/cm, and different temperatures from ambient to 220\textcelsius\ in the orthorombhic ER phase with some octahedral tilting.
The P-E curve area peaks at room temperature and decreases with temperature due to reduced residual polarization and coercive electric field. The hysteresis curve intersects the electric field axis at the coercive electric field and cuts on the residual polarization at polarization axis. Refer to Fig. \ref{fig:Comparison-of-P-E} (b) for the P-E characteristics curve at various temperatures and phases. At the reference temperature, the curve in R3c phase shows maximum residual polarization and P-E hysteresis region, suggesting good piezoelectric behaviour. The remnant polarization reduces to 18 $\mu C/m^{2}$ at 75\textcelsius, which was 30 $\mu C/m^{2}$ at reference temperature. When the temperature is raised to 75\textcelsius~from reference temperature, the coercive field drops from 85 to 70 $kV/cm$. Increasing temperature is expected to decrease the size of the hysteresis curve. At $T_{f}$, 190\textcelsius, the NR/ER phase boundary exists. BNT has strong residual polarization and high coercive electric fields, making it ideal for piezoelectric and high-temperature sensor applications.
After reaching this temperature, the NR/ER phase border emerges, causing the domain to transition to the ER phase relaxor, reducing the P-E hysteresis area and residual polarization. Negligible hysteresis area and residual polarization enable great strain response in the BNT, which will be studied further. The ER phase's hysteresis loop makes it ideal for high-temperature actuators, lead-free haptic devices, and human-computer interfaces.
The development of P-E characteristics with temperature allows us to examine BNT's versatility and promise in sustainable technology applications. BNT's complicated microdynamics enable the creation of several eco-friendly, sustainable technical solutions for various operational situations.
\vspace{-0.25cm}
\section{Conclusions}
\vspace{-0.25cm}
Experimentally verified phase-field thermo-electromechanical computational model explored BNT ceramics' phase change and micro-domain switching behaviour. The model predicts BNT ceramics' complicated multi-phase behaviour. Some related study findings are as follows:
(a) The maximal P-E curve area occurs at reference temperature with strong residual polarization and coercive electric field, but decreases with temperature progression,
(b) BNT ceramics have significant piezoelectric response up to the NR/ER phase boundary, making them an eco-friendly option for high-temperature haptics and sensors, and
(c) Since the ER phase has low residual polarization at this temperature range, the hysteresis curve area diminishes, giving it a viable candidate for sustainable high-temperature actuator applications.

The present study fully captures phase change and micro-domain switching. The P-E characteristics development with temperature allows us to examine BNT's piezoelectric behaviour at various temperatures, which helps us comprehend its use in high-temperature, sensor, actuator, and haptic device applications. Computational modelling, which is cheaper than experiments, supports these conclusions. We expect this study will contribute to establishing strong computational frameworks for predicting complicated material behaviours, providing innovative environmentally friendly solutions in sustainable technology.
\vspace{-0.25cm}
\subsection*{Acknowledgments:} 
The authors are grateful to the NSERC and the CRC Program (Canada) for their support. This publication is part of the R$^{+}$D$^{+}$i project, PID2022-137903OB-I00, funded by MICIU/AEI/10.13039/ 501100011033/ and by FEDER, EU. This research was enabled in part by support provided by SHARCNET (www.sharcnet.ca) and Digital Research Alliance of Canada (www.alliancecan.ca).
\vspace{-0.25cm}
\subsection*{Disclosure of conflict of interests} 
We state that "No Competing interests are at stake and there is No Conflict of Interest with other people or organizations".
%
%
%

\end{document}